\documentclass[a4paper, 11pt]{article}
\usepackage[pdftex, a4paper]{geometry}
\usepackage{graphicx}				
				
\usepackage{amsmath}
\usepackage{amsfonts}
\usepackage{dsfont}
\usepackage{mathrsfs}
\usepackage{amssymb}
\usepackage{bbm}
\usepackage{epsfig}
\usepackage[english]{babel}
\usepackage[all]{xy}

\newtheorem{theorem}{Theorem}
\newtheorem{definition}{Definition}

\newtheorem{lemma}{Lemma}

\newtheorem{proposition}{Proposition}

\begin{document}

\title{\bf Forgetting  in the  Synchronization of Quantum Networks}
\date{}

\author{ Shuangshuang Fu, Guodong Shi,  and  Ian R. Petersen\footnote{S. Fu and G. Shi  are with the Research School of Engineering, The Australian National University, Canberra, Australia. I. R. Petersen is with School of Engineering and Information Technology, University of New South Wales, Canberra, Australia. Email: shuangshuang.fu, guodong.shi@anu.edu.au,  i.r.petersen@gmail.com.}}
\maketitle

\begin{abstract}
In this paper, we study  the decoherence property of synchronization master equation for  networks of qubits interconnected  by swapping operators. The network Hamiltonian is assumed to be diagonal with different  entries so that it might  not be commutative with the swapping operators. We prove a theorem establishing a general  condition under which almost complete decohernece is achieved, i.e., all but two of  the off-diagonal entries of the network density operator asymptotically tend to zero. This result explicitly shows that   quantum dissipation networks tend to forget the information initially encoded when the internal (induced by network Hamiltonian) and external (induced by swapping operators) qubit interactions do not comply with each other.
\end{abstract}


\section{Introduction}
Inspired by the  developments of distributed consensus control for classical network systems in the past decade \cite{tsi,jad03,saber04}, consensus and synchronization problems of quantum networks have also recently attracted attention in the research community \cite{Sepulchre-non-communtative,Ticozzi,Ticozzi-SIAM,ShiDongPetersenJohansson}.  Sepulchre \emph{et al.}  \cite{Sepulchre-non-communtative} generalized consensus algorithms to non-commutative spaces
 and presented convergence results for quantum stochastic maps, and  showed how the Birkhoff theorem can be used to analyze the asymptotic convergence
 of a quantum system to a fully mixed state.  Mazzarella \emph{et al.} \cite{Ticozzi} made a systematic study regarding   consensus-seeking in quantum networks,
 introducing several classes of consensus quantum states and a quantum generalization to the gossip iteration
 algorithm  based on pairwise swapping operators for reaching a symmetric state (consensus) over a  quantum network.  The class of quantum gossip algorithms was further extended to
 symmetrization problems in a group-theoretic framework  \cite{Ticozzi-SIAM}.

The analysis of quantum consensus seeking was  further developed  using the  graphical methods  for studying classical network systems \cite{Magnus}, and it was shown that
the vectorized density operator evolving along  quantum consensus dynamics is equivalent to  a number of  parallel classical consensus dynamics over disjoint subgraphs \cite{ShiDongPetersenJohansson}, which  enabled us to study quantum consensus dynamics via their classical analogous with all details inherited. Furthermore, when the network Hamiltonian is commutative with the swapping operators, one can derive a so-called quantum synchronization master equation \cite{ShiDongPetersenJohansson} as the quantum counter part of the classical linear synchronization results \cite{WU-CHUA95,WUbook}. This quantum synchronization master equation can be physically realized via quantum dissipation networks where quantum nodes are interconnected by local environments \cite{NaturePhysiscs}.

It however has been understood that when a quantum system interacts with the environment through dissipative couplings, the quantum information encoded in  the system is often  washed out in the sense that the off-diagonal entries of the system density operator asymptotically vanish. This phenomenon is known as decoherence.  In this paper, we study  the decoherence property of synchronization master equation for quantum networks of qubits interconnected  by swapping operators. The network Hamiltonian is assumed to be diagonal but with different diagonal entries so it might  not be commutative with the swapping operators. We prove a theorem establishing a general  condition under which almost complete decohernece is achieved, i.e.,  all but two of  the off-diagonal entries of the network density operator asymptotically tend to zero. This result explicitly shows that   quantum dissipation networks tend to forget the information initially encoded when the internal  and external  qubit interactions, respectively induced by the network Hamiltonian and the swapping operators,  do not comply with each other.

The remainder of the paper is organized as follows. Section~\ref{Sec2} introduces the problem definition and presents the main result obtained. A brief introduction to the quantum mechanics related to the developments of the current paper is also provided in Section~\ref{Sec2} as well as a few numerical verifications of the theoretical result. Section~\ref{sec3} establishes the detailed proof of the main result, where the arguments are organized into step-by-step blocks. Finally Section \ref{Sec4} concludes the paper.
\section{Problem Definition,  Main Result, and  Examples}\label{Sec2}
In this section, we define the problem of interest, present the main result, and provide numerical examples illustrating the obtained result.
\subsection{Quantum States, Density Operators, and Partial Trace}
We first give a brief introduction to quantum systems' states. We refer the readers  to \cite{Nielsen} for a comprehensive treatment.

\subsubsection{Quantum States}
  The state space associated with any isolated quantum system is a complex vector space with inner product, {i.e.}, a Hilbert space $\mathcal{H}$. The system is completely described by its state vector, which is a unit vector in the system's state space and often denoted by $|\psi\rangle\in\mathcal{H}$  (known as the Dirac notion). The state space of a composite quantum system is the tensor product of the state space of each component system, e.g., two quantum systems with state spaces $\mathcal{H}_A$ and $\mathcal{H}_B$, respectively, form a composite system with state space $\mathcal{H}_A \otimes \mathcal{H}_B$, where $\otimes$ stands for tensor product. If the two quantum systems are isolated respectively with states  $|\psi_A\rangle\in\mathcal{H}_A$ and $|\psi_B\rangle\in\mathcal{H}_B$, the composite system admits a state $|\psi_A\rangle\otimes |\psi_B\rangle$.

\subsubsection{Density Operators}
For an open quantum system, its state can also be described by a positive (i.e., positive semi-definite) Hermitian density operator  $\rho$ satisfying $\text{tr}(\rho)=1$. A quantum state $|\psi\rangle\in\mathcal{H}$, induces a linear operator, denoted $|\psi\rangle\langle\psi |$, by
\begin{align*}
  |\psi\rangle\langle\psi | \Big(|x\rangle\Big)=  \Big(|\psi \rangle, |x\rangle\Big)  |\psi\rangle
  \end{align*}
  with $\Big(\cdot,\cdot\Big)$ being the inner product\footnote{Under Dirac notion this inner product is written as $\Big(|\psi \rangle, |x\rangle\Big)=\langle\psi |x\rangle$, where $\langle\psi |$ is the dual vector of $|\psi \rangle$.} equipped by the Hilbert space $\mathcal{H}$. Then $\rho=|\psi\rangle\langle\psi |$ defines the corresponding density operator. Density operators provide a convenient description of {\it ensembles of pure state}: If a quantum system is in state $|\psi_i \rangle$ with probability $p_i$ where $\sum_i p_i=1$, its density operator is
  \begin{align*}
  \rho=\sum_i p_i   |\psi_i\rangle\langle\psi_i |.
  \end{align*}
Any  positive and Hermitian operator  with trace one defines a proper density operator describing certain quantum state, and vice versa.

\subsubsection{Partial Trace}
Let $\mathcal{H}_A$ and $\mathcal{H}_B$ be the state spaces of  two quantum systems $A$ and $B$, respectively. Their composite system is described by
a density operator $\rho^{AB}$. Let $\mathfrak{L}_A$, $\mathfrak{L}_B$, and  $\mathfrak{L}_{AB}$ be the spaces of (linear) operators over  $\mathcal{H}_A$,
 $\mathcal{H}_B$, and  $\mathcal{H}_A\otimes\mathcal{H}_B$, respectively.   Then the partial trace over system $B$, denoted by ${\rm Tr}_{\mathcal{H}_B}$,
 is an operator mapping  $\mathfrak{L}_{AB}$ to $\mathfrak{L}_{A}$ defined by
$$
{\rm Tr}_{\mathcal{H}_B}\Big(|p_A\rangle  \langle q_A| \otimes  |p_B\rangle  \langle q_B| \Big)= |p_A\rangle  \langle q_A|  {\rm Tr} \Big(  |p_B\rangle  \langle q_B| \Big)
$$
for all $|p_A\rangle, |q_A \rangle\in \mathcal{H}_A, |p_B\rangle, |q_B\rangle \in \mathcal{H}_B$.

The reduced density operator (state) for system $A$, when the composite system is in the state  $\rho^{AB}$, is defined as $\rho^A= {\rm Tr}_{\mathcal{H}_B}(\rho^{AB})$.
The physical interpretation of $\rho^A$ is that $\rho^A$ holds the full information of system $A$ in $\rho^{AB}$.

\subsection{Qubit  Network and Its Synchronization}
In quantum  systems, the two-dimensional Hilbert space  forms the state-space of {\it qubits} (the most basic quantum system).  Let $\mathcal{H}$ be a two-dimensional Hilbert space for qubits.  The standard  computational basis of $\mathcal{H}$ is denote by $|0\rangle$ and $|1\rangle$.
 An $n$-qubits quantum network is the composite quantum system of $n$ qubits in the set $\mathsf{V}=\{1,\dots,n\}$, whose state space is
  the Hilbert space $\mathcal{H}^{\otimes n}=\mathcal{H}\otimes \dots \otimes \mathcal{H}$, where $\otimes$ denotes the tensor product.
   The {\it swapping operator} between qubits $i$ and $j$, denoted as $U_{ij}$,  is defined by
\begin{align*}
&{U_{ij}} \big(|q_{1}\rangle \otimes \dots \otimes| q_{i}\rangle \otimes \dots \otimes |q_j\rangle \otimes \dots \otimes |q_n\rangle \big)= \nonumber\\
 &|q_{1}\rangle \otimes \dots \otimes |q_{j}\rangle\otimes \dots \otimes |q_i\rangle\otimes \dots \otimes |q_n\rangle,
\end{align*}
for all $q_i\in \{0,1\}, i=1,\dots,n$. In other words, the swapping operator $U_{ij}$ switches the information held in qubits $i$ and $j$
without changing the states of other qubits.

The density operator of  the $n$-qubit network is denoted as  $\rho$. A quantum interaction graph over the $n$-qubit  network is  an undirected, connected graph $\mathsf{G}=(\mathsf{V}, \mathsf{E})$, where  each element in $\mathsf{E}$,
called a quantum edge,  is an unordered pair of two distinct qubits  denoted as $\{i,j\}\in\mathsf{E}$ with $i,j\in\mathsf{V}$.  The  state evolution of the quantum network is given by the following     master equation \cite{ShiDongPetersenJohansson},
\begin{align}\label{sysLind}
\frac{d \rho}{dt}=-\frac{\imath}{\hbar}[H, \rho]+\sum_{\{j,k\}\in \mathsf{E}}  \Big(U_{jk}\rho U_{jk}^\dag  -\rho\Big),
\end{align}
where  $[\cdot,\cdot]$ denotes the commutator of two operators, $H$ is the effective Hamiltonian as a Hermitian operator over the underlying Hilbert space, $\imath^2=-1$, $\hbar$ is the reduced Planck constant,  $U_{jk}$ is the swapping operator between $j$ and $k$. As discussed in \cite{ShiDongPetersenJohansson}, the above synchronization dynamics is a Markovian master equation in the Lindblad form
 \cite{Lindblad 1976,Breuer and Petruccione 2002} and can be physically realized via building suitable local environments among the qubits \cite{NaturePhysiscs}.

Let $\mathbf{P}$ be the $n$'th permutation group and assume the initial time is $0$ for the system (\ref{sysLind}). It has been shown in \cite{ShiDongPetersenJohansson} that when the network Hamiltonian $H$ is commutative with the swapping operators, i.e., $[H,U_{jk}]=0$ for all $\{j,k\}\in \mathsf{E}$, quantum synchronization is achieved in the sense that (cf., \cite{ShiDongPetersenJohansson})
\begin{align}
\lim_{t\to \infty} \Big(\rho(t)-e^{-\imath Ht/\hbar} \rho_\ast e^{\imath H t/\hbar} \Big)=0
\end{align}
along the system (\ref{sysLind}), where $\rho_\ast=\frac{1}{n!} \sum_{\pi \in \mathbf{P}} U_\pi \rho(0) U^\dag_\pi$. Let $\rho^k(t):={\rm Tr}_{\otimes_{j\neq k} \mathcal{H}_j } (\rho(t))$ be the reduced state of qubit $k$ at time $t$. For the limiting trajectory, there holds for all $j=1,\dots,n$ that
\begin{align}
{\rm Tr}_{\otimes_{j\neq k} \mathcal{H}_j } \Big(e^{-\imath Ht/\hbar} \rho_\ast e^{\imath H t/\hbar}\Big)={\rm Tr}_{\otimes_{j=1}^{n-1} \mathcal{H}_j } \Big(e^{-\imath Ht/\hbar} \rho_\ast e^{\imath H t/\hbar}\Big),
\end{align}
which in turn leads to
\begin{align}
\lim_{t\to \infty} \Big(\rho^k(t)-\rho^m(t)\Big)=0, \ k,m\in\mathsf{V}.
\end{align}

\subsection{Main Result: A Quantum Forgetting Theorem}

Throughout the remainder   of the paper,  we investigate $\rho(t)$ under the following standard basis of $\mathcal{H}^{\otimes n}$:
\begin{align*}
\mathbb{B}:=\Big\{ |q_1 \dots q_n\rangle \langle p_1 \dots p_n|: p_i,q_i\in\{0,1\}, i\in\mathsf{V}\Big\}.
\end{align*}
We identify the operators with their matrix representations under the basis $\mathbb{B}$, for the ease of presentation. We denote  $\big[\rho(t) \big]_{|q_1 \dots q_n\rangle \langle p_1 \dots p_n|}$ as the $|q_1 \dots q_n\rangle \langle p_1 \dots p_n|$-entry of the density operator $\rho(t)$ under the basis $\mathbb{B}$.   The diagonal entries of the elements in $\mathbb{B}$ are put in the set
\begin{align*}
\mathbb{B}_{\rm D}:=\Big\{ |p_1 \dots p_n\rangle \langle p_1 \dots p_n|: p_i\in\{0,1\}, i\in\mathsf{V}\Big\}.
\end{align*}

For simplicity we always write $z=z_1 \dots z_n$ with  $z_i\in\{0,1\}, i\in\mathsf{V}$. In this paper, we are interested in the evolution of the system (\ref{sysLind}) in the absence of the commuting condition between the Hamiltonian and the swapping operators. Particularly, we are interested in the {\it decoherence} of  the system (\ref{sysLind}), i.e., decaying of the off-diagonal entries of the density operators. To be precise, we introduce the definition of decoherence in the following.

\begin{definition}
The system (\ref{sysLind}) achieves $|x\rangle \langle y|$-decoherence for if
$\lim_{t \to \infty} \big[\rho(t)\big]_{|x\rangle \langle y|}=0$.
\end{definition}

We impose a standing assumption on the network Hamiltonian $H$.

\medskip

\noindent {\bf Assumption} {\it There are $2^n$ real numbers $\lambda_{|p\rangle \langle p|}\in \mathds{R}$, $|p\rangle \langle p|\in\mathbb{B}_{\rm D}$ such that $H=\sum_{|p\rangle \langle p|\in\mathbb{B}_{\rm D}}\lambda_{|p\rangle \langle p|} |p\rangle \langle p|$.}

\medskip

Under the above assumption, the network Hamiltonian $H$ is diagonal under the standard basis. Since the Hamiltonian $H$ is a Hermitian operator, one can always find a basis  of $\mathcal{H}^{\otimes n}$ under which $H$ is represented by a diagonal matrix. The assumption that  $H$ is diagonal under the standard basis is however quite restrictive. Nevertheless, this assumption allows for basic non-commuting properties   between $H$ and the swapping operators,  and in the meantime enables  us  to derive some explicit result  for the decoherence of  the system (\ref{sysLind}).

Let ${\mathbf{C}_n^k}$ be the  combinatorial number of selecting $k$ from $n$ objectives. Denote $\mathbf{0}=0\dots0$ and $\mathbf{1}= 1\dots1$ both with $n$ digits.
 The following is our main result.

\medskip

\begin{theorem}\label{theoremforgetting}
The following statements hold for the system (\ref{sysLind}).
\begin{itemize}
\item[(i)]  If the elements $
\lambda_{|p\rangle \langle p|}-\lambda_{|p'\rangle \langle p'|}, \ \ p\neq p'
$
are pairwise distinct, then almost complete decoherence is achieved in the sense that $|x\rangle \langle y|$-decoherence is reached for all $x\neq y$ satisfying either  $x\notin\{\mathbf{0}, \mathbf{1}\}$ or $y\notin\{\mathbf{0}, \mathbf{1}\}$, for which the convergence is at an exponential rate;

\item[(ii)] $[\rho(t)]_{|\mathbf{0}\rangle\langle \mathbf{1}|}=[\rho(0)]_{|\mathbf{0}\rangle\langle \mathbf{1}|}e^{-{\imath}( \lambda_{|\mathbf{0}\rangle \langle \mathbf{0}|}-\lambda_{|\mathbf{1}\rangle \langle \mathbf{1}|})t/{\hbar}}$;   $[\rho(t)]_{|\mathbf{1}\rangle\langle \mathbf{0}|}=[\rho(0)]_{|\mathbf{1}\rangle\langle \mathbf{0}|}e^{-{\imath}( \lambda_{|\mathbf{1}\rangle \langle \mathbf{1}|}-\lambda_{|\mathbf{0}\rangle \langle \mathbf{0}|})t/{\hbar}}$;

\item[(iii)] For $x=x_1\dots x_n$ with $\sum_{i=1}^n x_i =k$, there holds that
$$
\lim_{t \to \infty} \big[\rho(t)\big]_{|x\rangle \langle x|}={\sum_{y:\sum_{i=1}^n y_i =k}} \big[\rho(0)\big]_{|y\rangle \langle y|}/{\mathbf{C}_n^k}
$$
where the convergence is also exponential. Consequently, there are at most $n+1$ different values for the limits of the diagonal entries of $\rho(t)$.

\end{itemize}
\end{theorem}
\medskip

Note that there are only two off-diagonal entries, $[\rho(t)]_{|\mathbf{0}\rangle\langle \mathbf{1}|}$ and  $[\rho(t)]_{|\mathbf{1}\rangle\langle \mathbf{0}|}$, that can possibly  be not vanishing   for the system (\ref{sysLind}).  It is clear that if the elements $
\lambda_{|p\rangle \langle p|}-\lambda_{|p'\rangle \langle p'|}, \ \ p\neq p'
$ are pairwise distinct, the network Hamiltonian is no longer commutative with the swapping operators in the system (\ref{sysLind}). The  decoherence result established Theorem \ref{theoremforgetting} reveals that the quantum network along the system (\ref{sysLind}) then tends to {\it forget} almost all the information contained in the off-diagonal entries of the initial value $\rho(0)$, which represent initial  correlations among  the the basis states \cite{Nielsen}.  Furthermore, if the condition that the elements $
\lambda_{|p\rangle \langle p|}-\lambda_{|p'\rangle \langle p'|}, \ \ p\neq p'
$ are  pairwise distinct does not hold strictly, it is clear from the proof of Theorem \ref{theoremforgetting} that  a network Hamiltonian $H$ being non-commutative with the swapping operators in the system (\ref{sysLind}) continues to tend to wash out the off-diagonal entries of the network density operator, just possibly leaving a few nonzero off-diagonal entries.

\subsection{Numerical Example}
In this subsection, we present a numerical example to illustrate the obtained main result. We consider three qubits indexed in $\mathsf{V}=\{1,2,3\}$. Their interaction graph is  fixed as the complete graph, i.e., $\mathsf{E}=\big\{\{1,2\}, \{2,3\}, \{1,3\} \big\}$.
Let $\alpha_{12}=\alpha_{13}=\alpha_{23}=1$. We denote $x=x_1x_2x_3$ with $x_i\in\{0,1\}$, and whenever applicable we identify $x$ as a binary number.   The initial network state is chosen to be
$$
\rho_0=\Big( \sum_{x}|x\rangle  \Big)\Big( \sum_{x}\langle x|  \Big)/128+ \Big( \sum_{|x\rangle \langle x|\in\mathbb{B}_{\rm D}} (x+1) |x\rangle \langle x| \Big)/72.
$$
The network Hamiltonian is chosen to be $$
H=\sum_{|x\rangle \langle x|\in\mathbb{B}_{\rm D}} 2^{x} |x\rangle \langle x|
$$
so that our standing assumption is satisfied.

We first plot the evolution of $ \big[\rho(t)\big]_{|x\rangle \langle x|}$ for all $x$.  Clearly the eight trajectories  are asymptotically grouped into four clusters. We also introduce
\begin{align}
E_o(t):= \sum_{|x\rangle \langle y|: \ x\neq y,  |x\rangle \langle y|\notin \{|\mathbf{0}\rangle \langle 1|,|\mathbf{1}\rangle \langle 0|\}} \Big\| \big[\rho(t)\big]_{|x\rangle \langle y|}\Big\|^2
\end{align}
as a measure of decoherence for all off-diagonal entries of $\rho(t)$ except for $|\mathbf{0}\rangle \langle 1|$ and $|\mathbf{1}\rangle \langle 0|$. We also plot $E_o(t)$ and clearly it tends to zero exponentially.

\begin{figure}
\begin{minipage}[t]{0.5\linewidth}
\centering
\includegraphics[width=3.0in]{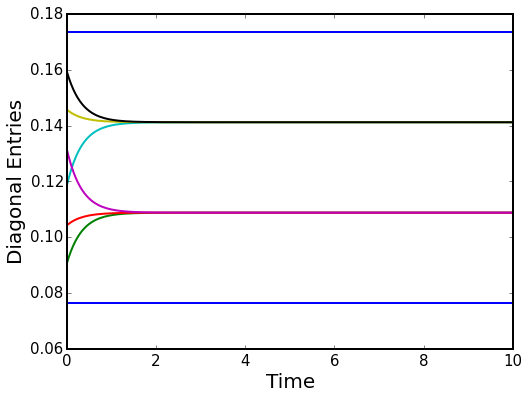}
\end{minipage}%
\begin{minipage}[t]{0.5\linewidth}
\centering
\includegraphics[width=3.0in]{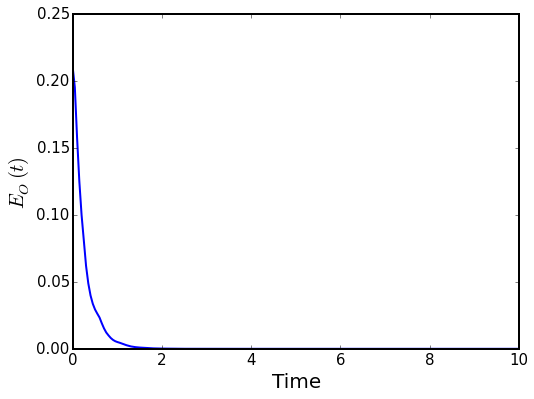}
\end{minipage}
\caption{The evolution of the diagonal entries of the network density operator (left) and the $E_o(t)$ (right).}
\label{fig:bloch}
\end{figure}

\section{Proof of the Main Result}\label{sec3}
In this section, we present the proof of Theorem \ref{theoremforgetting}. The analysis is based on splitting   the entries of the density operator into decoupled subgroups where interactions only take place  inside each subgroup. The idea of breaking down large density operators of multiple qubits can in fact be traced back to \cite{Altafini} using Stokes tensors. In \cite{ShiDongPetersenJohansson}, the method of investigating  the interconnection of the individual entries  of the network density operator was systematically studied.

\subsection{Graphical Decomposition}

We first establish a technical lemma.

\medskip

\begin{lemma}\label{lem1}
Denote $\mathrm{C}$ as an operator over the space of $\rho$ by $\mathrm{C}(\rho):=[H, \rho]$. Then $[\mathrm{C}]_{|x\rangle \langle y|}=\big( \lambda_{|x\rangle \langle x|}-\lambda_{|y\rangle \langle y|}\big)|x\rangle \langle y|$.
\end{lemma}
{\it Proof.} Based on our standing assumption on the definition of $H$, we obtain
\begin{align}
[\mathrm{C}]_{|x\rangle \langle y|}&=\sum_{|p\rangle \langle p|\in\mathbb{B}_{\rm D}}\lambda_{|p\rangle \langle p|} |p\rangle \langle p|x\rangle \langle y|-\sum_{|p\rangle \langle p|\in\mathbb{B}_{\rm D}}\lambda_{|p\rangle \langle p|}|x\rangle \langle y |p\rangle \langle p|\nonumber\\
&=\sum_{|p\rangle \langle p|\in\mathbb{B}_{\rm D}}\lambda_{|p\rangle \langle p|} \delta(p,x) |p\rangle \langle y|-\sum_{|p\rangle \langle p|\in\mathbb{B}_{\rm D}}\lambda_{|p\rangle \langle p|} \delta(p,y)|x\rangle  \langle p|\nonumber\\
&=\big( \lambda_{|x\rangle \langle x|}-\lambda_{|y\rangle \langle y|}\big)|x\rangle \langle y|,
\end{align}
where $\delta(a,b)=1$ if $a=b$ and $\delta(a,b)=0$ otherwise. This completes the proof. \hfill$\square$

\medskip

We also recall the following lemma, which is a variation of the Lemma 4  in \cite{ShiDongPetersenJohansson}.

\medskip
\begin{lemma}\label{lem2}
Let $u_{ij}$ in the swapping between $i$ and $j$ in the permutation group $\mathbf{P}$, i.e., $u_{ij}(i)=j$, $u_{ij}(j)=i$, and $u_{ij}(k)=k$ for $k\neq i,j$. Then there holds $U_{jk}{|x\rangle \langle y|} U_{jk}^\dag=|u_{jk}(x)\rangle \langle u_{jk}(y)|$.
\end{lemma}
\medskip

In light of Lemmas \ref{lem1} and \ref{lem2}, we can now rewrite the system (\ref{sysLind}) into its entry-wise equivalence:
\begin{align}\label{syscomponent}
\frac{d}{dt} \big[\rho(t) \big]_{|x\rangle \langle y|}= -\frac{\imath}{\hbar}\big( \lambda_{|x\rangle \langle x|}-\lambda_{|y\rangle \langle y|}\big)\big[\rho(t) \big]_{|x\rangle \langle y|}+\sum_{\{j,k\}\in \mathsf{E}}  \Big(\big[\rho(t) \big]_{|u_{jk}(x)\rangle \langle u_{jk}(y)|}  - \big[\rho(t) \big]_{|x\rangle \langle y|}\Big),
\end{align}
where $x=x_1\dots x_n$, $y=y_1\dots y_n$ with $x_i,y_i\in\{0,1\}$, $i\in\mathsf{V}$. We see from (\ref{syscomponent}) that (cf., Lemma 5 \cite{ShiDongPetersenJohansson})
$$
\mathcal{R}_{|x_1 \dots x_n\rangle \langle y_1 \dots y_n|}:=\Big\{ |x_{\pi(1)} \dots x_{\pi(n)}\rangle \langle y_{\pi(1)} \dots y_{\pi(n)}|:\pi\in \mathbf{P} \Big\}
$$
forms a subset of entries whose state evolution is not influenced by entries outside.
It is clear that if either  $x\notin\{\mathbf{0}, \mathbf{1}\}$ or $y\notin\{\mathbf{0}, \mathbf{1}\}$ holds, then $|\mathcal{R}_{|x\rangle \langle y|}|\geq2$ from Lemma \ref{lem5} of \cite{ShiDongPetersenJohansson}. Moreover, Theorem \ref{theoremforgetting}.(ii) follows from direct calculation from the system (\ref{syscomponent}) since $\mathcal{R}_{|\mathbf{0}\rangle \langle \mathbf{1}|}$ and $\mathcal{R}_{|\mathbf{1}\rangle \langle \mathbf{0}|}$ are singletons.

 We are now ready to state the following lemma which transforms the  decoherence of system (\ref{sysLind}) to a synchronization problem of a classical network.

\medskip

\begin{lemma}\label{lem3}
Consider a classical network with $N$ nodes indexed in the set $\mathrm{V}=\{1,\dots,N\}$ with an underlying interaction graph $\mathrm{G}=(\mathrm{V},\mathrm{E})$ which  is undirected and connected. Let node $i$ possess   a state $X_i \in \mathds{C}$. The evolution of the $X_i$ is given by
\begin{align}\label{sysmclassical}
\frac{d}{dt} X_i(t)=\imath \theta_i X_i(t)+\sum_{j:\{i,j\}\in\mathrm{E}} \big(X_j(t)-X_i(t) \big)
\end{align}
where $\theta_i \in \mathds{R}$ for all $i\in\mathrm{V}$. The following statements are equivalent.

\noindent (i) The system (\ref{sysmclassical}) satisfies $\lim_{t\to \infty} X_i(t)=0$ for all $i\in\mathrm{V}$ if $N\geq2$ and  the $\theta_i$, $i\in\mathrm{V}$ are pairwise distinct;

\noindent (ii) The system (\ref{sysLind}) achieves $|x\rangle \langle y|$-decoherence  if $|\mathcal{R}_{|x\rangle \langle y|}|\geq2$ and the $
\lambda_{|p\rangle \langle p|}-\lambda_{|p'\rangle \langle p'|}$, $p\neq p'
$
are pairwise distinct.
\end{lemma}
{\it Proof.}  Denoting $\theta_i=-\big( \lambda_{|x\rangle \langle  x|}-\lambda_{|y\rangle \langle y|}\big)/{\hbar}$ and investigating the system (\ref{syscomponent}) over the set $\mathcal{R}_{|x\rangle \langle y|}$ with $x\neq y$, the desired equivalence becomes clear from the definition of  decoherence immediately. \hfill$\square$

\subsection{A Classical Detour}
We proceed to make  a further investigation to the system (\ref{sysmclassical}). To this end, we make use of the realification method to investigate  the system (\ref{sysmclassical}) via studying  the real and imaginary parts separately. We write
\begin{align*}
X_i(t)=R_i(t)+ \imath S_i(t)
\end{align*}
where $R_i(t)$ and $S_i(t)$ are the real and imaginary components of $X_i(t)$, respectively. Denote $Y_i(t)=(R_i(t)\ S_i(t))^T$ for $i\in\mathrm{V}$. Then the system (\ref{sysmclassical}) reads
\begin{align}\label{sysreal}
\frac{d}{dt} Y_i(t) =A_i Y_i(t)+\sum_{j:\{i,j\}\in\mathrm{E}} \big(Y_j(t)-Y_i(t) \big), \ \ i\in\mathrm{V}
\end{align}
where
\begin{align*}
A_i=\begin{pmatrix}
  0 & -\theta_i \\
  \theta_i & 0
 \end{pmatrix},\ \ i\in\mathrm{V}.
\end{align*}
Clearly the system (\ref{sysreal}) defines  a classical  linear synchronization problem with non-identical node self dynamics specified by the $A_i$ (cf., \cite{WU-CHUA95,WUbook}). The following is an intermediate result for  the  system (\ref{sysmclassical}) established by studying its realification system (\ref{sysreal}).

\medskip

\begin{lemma} \label{lem4}
Denote $f(t):=\max_{i\in\mathrm{V}} \big\| X_i(t)\big\|^2$. Then $f(t)$ is a non-increasing function along the system (\ref{sysmclassical}).
\end{lemma}
{\it Proof.} Clearly $f(t)$ is a continuous but not necessarily continuously differentiable function. In this step, we prove that $f(t)$ is a non-increasing function along the system (\ref{sysreal}) by showing that its Dini derivative is always non-positive.

The upper Dini derivative of a function $h:(a,b)\rightarrow \mathds{R}$ at $t\in(a,b)$ is defined as \cite{Dini}
\begin{align}
D^+h(t)=\limsup_{s\rightarrow 0^+} \frac{h(t+s)-h(t)}{s}.
\end{align}
Define  $\mathcal{I}(t):=\arg \max_{i\in\mathrm{V}} \big\| Y_i(t)\big\|^2$. The  Lemma 2.2 of \cite{Lin07} enables us to derive
\begin{align}\label{1}
D^+ f(t)&= \max_{i\in \mathcal{I}(t)} \frac{d}{dt} \big\| Y_i(t)\big\|^2\nonumber\\
&= 2 \max_{i\in \mathcal{I}(t)} \Big \langle Y_i(t), A_i Y_i(t)+\sum_{j:\{i,j\}\in\mathrm{E}} \big(Y_j-Y_i \big)\Big\rangle \nonumber\\
&\stackrel{a)}{=}2 \max_{i\in \mathcal{I}(t)} \Big \langle Y_i(t), \sum_{j:\{i,j\}\in\mathrm{E}} \big(Y_j(t)-Y_i(t) \big)\Big\rangle \nonumber\\
&\stackrel{b)}{\leq} - \max_{i\in \mathcal{I}(t)}\sum_{j:\{i,j\}\in\mathrm{E}}   \Big(\big\|Y_i(t)\big\|^2 -\big\|Y_j(t)\big\|^2\Big) \nonumber\\
&\stackrel{c)}{\leq} 0,
\end{align}
where $a)$ is based on the fact that $\langle Y_i(t), A_i Y_i(t) \rangle=0$ from the definition of $A_i$, $b)$ follows from the elementary inequality $a^Tb\leq (\|a\|^2+\|b\|^2)/{2}$ for two vectors $a$ and $b$, and $c)$ is due to the definition of  $\mathcal{I}(t)$. Based on the properties of the Dini derivative, (\ref{1}) leads to that $f(t)$ is a non-increasing function along the system (\ref{sysreal}) for all $t\geq 0$. \hfill$\square$

\medskip

We are now ready to prove the following key lemma for the system (\ref{sysmclassical}).

\medskip

\begin{lemma}\label{lem5}
For the system (\ref{sysmclassical}) with initial value $X(0)=(X_1(0) \dots X_N(t))^T$, there exists a non-negative real number $Z_{X(0)}\geq 0$ such that $\lim_{t\to \infty} \big\|X_i(t)\big\|=Z_{X(0)}$ for all $i\in\mathrm{V}$.
\end{lemma}
{\it Proof.}  The analysis will be carried out for the system (\ref{sysreal}). Since $f(t)$ is a non-increasing by Lemma \ref{lem4}, for the initial value $X(0)=(X_1(0) \dots X_N(t))^T$, there exists  a constant $f^\ast({X(0)})\geq 0$ such that $\lim_{t\to \infty} f(t)=f^\ast$. We prove the desired lemma by showing  that $\lim_{t\to\infty} \big\| Y_i(t)\big\|^2=f^\ast$ for all $i\in\mathrm{V}$ via a contradiction argument.

Suppose there is a node $i_0\in\mathrm{V}$ satisfying $g_\ast:=\liminf_{t\to \infty}  \big\| Y_{i_0}(t)\big\|^2<f_\ast$. Consequently, there exists an infinite time sequence $t_1\leq \cdots < t_m<\cdots$ such that
\begin{align}\label{3}
 \big\| Y_{i_0}(t_m)\big\|^2 \leq \frac{1}{2}(g_\ast +f_\ast),\ \ m=1,2,\dots.
\end{align}
On the other hand, from the definition of $f_\ast$ and the analysis of Step 1 we conclude that for any $\epsilon>0$, there exists $T_\epsilon >0$ such that
\begin{align}
\big\| Y_{i}(t)\big\|^2 \leq f_\ast+\epsilon, \ \ t\geq T_\epsilon.
\end{align}

We build the remainder  of the proof in steps.

\medskip

\noindent Step 1. Take a time instant $t_m$ and without loss of generality let $t_m>T_\epsilon$. In this step, we bound $\big\| Y_{i_0}(t)\big\|^2$ during the time interval $[t_m, t_m+1]$. Similar to the derivation of (\ref{1}), we have \begin{align}
 \frac{d}{dt} \big\| Y_{i_0}(t)\big\|^2&{=}2 \Big \langle Y_{i_0}(t), \sum_{j:\{i_0,j\}\in\mathrm{E}} \big(Y_j(t)-Y_{i_0}(t) \big)\Big\rangle \nonumber\\
&{\leq} \sum_{j:\{i_0,j\}\in\mathrm{E}}  \Big(\big\|Y_j(t)\big\|^2 -\big\|Y_{i_0}(t)\big\|^2\Big) \nonumber\\
&{\leq} (n-1)\Big( f_\ast+\epsilon-\big\|Y_{i_0}(t)\big\|^2 \Big)
\end{align}
for all $t\geq T_\epsilon$. Invoking the Gr\"{o}nwall's inequality, we further conclude
\begin{align}\label{2}
\big\| Y_{i_0}(t)\big\|^2\leq e^{-(n-1)(t-t_m)}  \big\| Y_{i_0}(t_m)\big\|^2 +\Big(1- e^{-(n-1)(t-t_m)}\Big)\big(f_\ast+\epsilon\big), \ t\geq t_m.
\end{align}
Plugging in (\ref{3}), (\ref{2}) leads to
\begin{align}\label{4}
\big\| Y_{i_0}(t)\big\|^2\leq  \zeta g_\ast +\big(1- \zeta\big)\big(f_\ast+\epsilon\big), \ t\in[t_m, t_m+1],
\end{align}
where $\zeta= e^{-(n-1)}/2  $.

\medskip

\noindent Step 2. Now that the graph $\mathrm{G}$ is connected, there must be a node $i_1\neq i_0$ such that $\{i_0,i_1\}\in\mathrm{E}$. In this step, we bound $\big\| Y_{i_1}(t_m+1)\big\|^2$. For $\big\| Y_{i_1}(t)\big\|^2$, we have
\begin{align}
 \frac{d}{dt} \big\| Y_{i_1}(t)\big\|^2
&{\leq}  \Big(\big\|Y_{i_0}(t)\big\|^2 -\big\|Y_{i_1}(t)\big\|^2\Big) +\sum_{j\neq i_0:\{i_1,j\}\in\mathrm{E}}  \Big(\big\|Y_j(t)\big\|^2 -\big\|Y_{i_1}(t)\big\|^2\Big) \nonumber\\
&{\leq}  \zeta g_\ast +\big(1- \zeta\big)\big(f_\ast+\epsilon\big) -\big\|Y_{i_1}(t)\big\|^2+ (n-2)\Big( f_\ast+\epsilon-\big\|Y_{i_1}(t)\big\|^2 \Big)
\end{align}
for all $ t\in[t_m, t_m+1]$, where in the second inequality we have used (\ref{4}). Again,  invoking the Gr\"{o}nwall's inequality, we  conclude
\begin{align}\label{4}
\big\| Y_{i_1}(t_m+1)\big\|^2&\leq  e^{-(n-1)} \big\| Y_{i_1}(t_m)\big\|^2 +\big(1-e^{-(n-1)}\big)\Big( \zeta g_\ast +\big(1- \zeta\big)\big(f_\ast+\epsilon\big)+(n-2)( f_\ast+\epsilon)\Big)/(n-1)\nonumber\\
&\leq  e^{-(n-1)}( f_\ast+\epsilon) +\big(1-e^{-(n-1)}\big)\Big( \zeta g_\ast +\big(1- \zeta\big)\big(f_\ast+\epsilon\big)+(n-2)( f_\ast+\epsilon)\Big)/(n-1)\nonumber\\
&\leq (\phi \zeta) g_\ast+ \Big(1-(\phi \zeta)  \Big)( f_\ast+\epsilon),
\end{align}
where $\phi=(1-e^{-(n-1)})/(n-1) $. In fact, we even know
\begin{align}
\big\| Y_{s}(t_m+1)\big\|^2\leq (\phi \zeta) g_\ast+ \Big(1-(\phi \zeta)  \Big)( f_\ast+\epsilon), \ s\in\{i_0,i_1\}
\end{align}
since $\phi\in(0,1)$.

\medskip

\noindent Step 3. Since the graph $\mathrm{G}$ is connected, we can recursively apply the arguments in the Steps 2 and 3 to the rest of the nodes, and eventually establish
\begin{align}
\big\| Y_{s}(t_m+n-1)\big\|^2\leq (\phi \zeta)^{n-1} g_\ast+ \Big(1-(\phi \zeta)^{n-1}  \Big)( f_\ast+\epsilon), \ s\in \mathrm{V}.
\end{align}
This implies
\begin{align}
f(t_m+n-1)\leq (\phi \zeta)^{n-1} g_\ast+ \Big(1-(\phi \zeta)^{n-1}  \Big)( f_\ast+\epsilon),
\end{align}
which contradicts  the definition of $f_\ast$ if
\begin{align}
\epsilon<\frac{(\phi \zeta)^{n-1}(f_\ast- g_\ast)}{1-(\phi \zeta)^{n-1}}.
\end{align}
Therefore, we have proved that $\liminf_{t\to \infty}  \big\| Y_{i}(t)\big\|^2=f_\ast$ for all $i\in \mathrm{V}$. On the other hand, there always holds $\limsup_{t\to \infty}  \big\| Y_{i}(t)\big\|^2\leq f_\ast$ in light of Lemma \ref{lem4}. Consequently, we have shown that $\lim_{t\to \infty}  \big\| Y_{i}(t)\big\|^2=f_\ast$ for all $i\in\mathrm{V}$, which completes the proof. \hfill$\square$

\subsection{Completion of the Proof}
In this subsection, we complete the  proof of Theorem \ref{theoremforgetting}.
\subsubsection{Decoherence}
With Lemma \ref{lem3}, the  decoherence statement for the system (\ref{sysLind}) holds if $\lim_{t\to \infty} X_i(t)=0$ for all $i\in\mathrm{V}$ for the system (\ref{sysmclassical}) with $N\geq 2$ when  the $\theta_i$, $i\in\mathrm{V}$ are pairwise distinct. In fact, we are going to show a slightly stronger result  for the system  (\ref{sysmclassical}) which only requires that  there exist two distinct values within the $\theta_i$.

We recall a few preliminary results on the limiting set of autonomous systems.  Consider the following
autonomous system
\begin{equation}
\label{i1} \dot{x}=f(x),
\end{equation}
where $f:\mathds{R}^d\rightarrow \mathds{R}^d$ is a  continuous function. Let $x(t)$ be a solution of
(\ref{i1}) with initial condition $x(t_0)=x^0$. Then $\Omega_0\subset \mathds{R}^d$ is called a {\it positively invariant
set} of (\ref{i1}) if, for any $t_0\in\mathds{R}$ and any $x^0\in\Omega_0$,
we have $x(t)\in\Omega_0$, $t\geq t_0$, along  every solution $x(t)$ of (\ref{i1}).

We call $y$ a  $\omega$-limit point of $x(t)$ if there exists a  sequence $\{t_k\}$ with $\lim_{k\rightarrow \infty}t_k=\infty$ such that
$
\lim_{k\rightarrow \infty}x(t_k)=y.
$
The set of all $\omega$-limit points of $x(t)$ is called the  $\omega$-limit set of  $x(t)$,  and is denoted as $\Lambda^+\big(x(t)\big)$.
The following conclusion is well-known  \cite{rou}.
\begin{lemma}\label{leminvariant}
Let  $x(t)$ be a solution of (\ref{i1}). Then  $\Lambda^+\big(x(t)\big)$ is positively  invariant. Moreover, if $x(t)$ is contained in a compact set,
then $\Lambda^+\big(x(t)\big)\neq \emptyset$.
\end{lemma}

We are now ready to state the following result for the system (\ref{sysmclassical}).

\medskip

\begin{figure}[t]
\begin{center}
\includegraphics[height=2.8in]{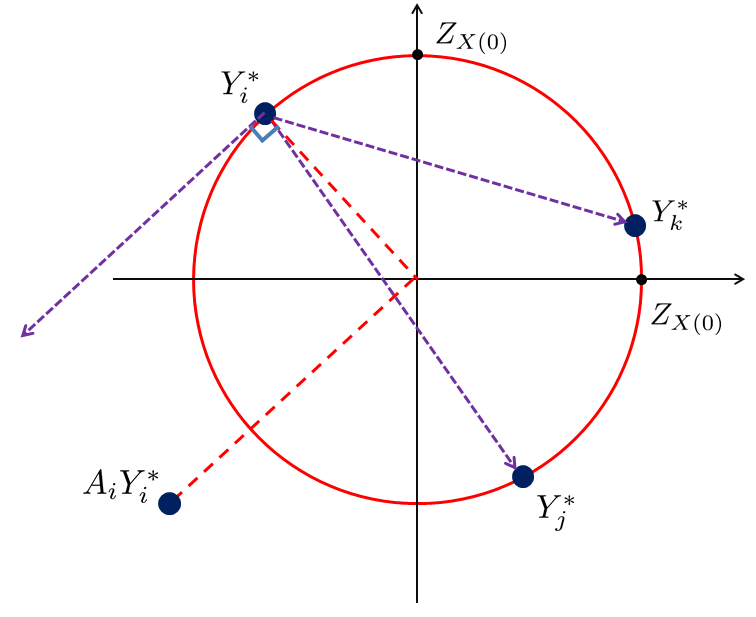}
\caption{Illustration to why  there must hold $Y_1^\ast=\dots=Y_N^\ast$ for any $Y_\ast\in\Lambda^+\big(Y(t)\big)$: (i) If $Y_i^\ast\neq Y_j^\ast$ and there is a link between $i$ and $j$, the trajectory starting from $Y_\ast$ must leave  $\Delta$ since the terms  $A_i Y_i^\ast$ are always perpendicular to the tangential directions of the manifold $\Delta$; (ii) such a pair always exists if there are $i,j\in\mathrm{V}$ with $Y_i^\ast\neq Y_j^\ast$ since the graph $\mathrm{G}$ is connected. }
\label{invariant}
\end{center}
\end{figure}

\begin{proposition}\label{prop1}
For  the system (\ref{sysmclassical}) with $N\geq 2$, the following statements hold.

\noindent (i) $\lim_{t\to \infty} X_i(t)=0$ for all $i\in\mathrm{V}$ if   there exist at least two distinct values within the $\theta_i$.

\noindent (ii) $\lim_{t\to \infty} \big\| X_i(t)-e^{\imath \theta t} \frac{\sum_{i=1}^N X_i(0)}{N}\big\|=0$  if there is $\theta\in \mathds{R}$ such that $\theta=\theta_i$ for all $i\in\mathrm{V}$.
\end{proposition}
{\it Proof.} (i) By Lemma \ref{lem4}, for any given initial value, the trajectory of  the realification system (\ref{sysreal}) is contained in a compact set. Therefore, the  $\omega$-limit set of  $Y(t)=(Y_1(t) \dots Y_N(t))^T$ along the   system (\ref{sysreal}) is nonempty and invariant in light of Lemma \ref{leminvariant}. On the other hand, making use of the Lemma \ref{lem5}, one finds that  the  $\omega$-limit set of $Y(t)$, denoted $\Lambda^+\big(Y(t)\big)$, can only be a subset of the set
\begin{align}
\Delta:=\Big\{Y^\ast=(Y_1^\ast \dots Y_N^\ast): \|Y_i^\ast\|=Z_{X(0)} \Big\}.
\end{align}
Without loss of generality we assume $Z_{X(0)}>0$ since otherwise the desired result holds immediately. The remaining argument relies on showing  that any subset of $\Delta$ cannot be invariant for the   system (\ref{sysreal}) if   there exist at least two distinct values within the $\theta_i$. We only need to establish  two facts.
\begin{itemize}
\item[F1)] For any $Y_\ast\in\Lambda^+\big(Y(t)\big)$, there must hold $Y_1^\ast=\dots=Y_N^\ast$. This is due to that as long as $Y_i^\ast\neq Y_j^\ast$, the trajectory starting from $Y_\ast$ must leave the set $\Delta$ since the terms  $A_i Y_i^\ast$ are always perpendicular to the tangential directions of the manifold $\Delta$. See Figure \ref{invariant} for an illustration.
    \item[F2)] From F1, we have $\Lambda^+\big(Y(t)\big)\subseteq \Delta \bigcap \Upsilon$ with $\Upsilon:=\Big\{Y^\ast=(Y_1^\ast \dots Y_N^\ast): Y_1^\ast=\dots=Y_N^\ast \Big\}$. However, $\Upsilon$ cannot be invariant if there are at least two distinct values within the $\theta_i$ since $Z_{X(0)}>0$.
\end{itemize}

Therefore, one must have  $Z_{X(0)}=0$ and the desired conclusion follows.

\medskip

\noindent (ii) The conclusion is straightforward using the transformation $\tilde{X}_i(t)=e^{-\imath \theta t} X_i(t)$, where clearly
$$
\frac{d}{dt}\tilde{X}_i(t)=\sum_{j:\{i,j\}\in\mathrm{E}} \big(\tilde{X}_j(t)-\tilde{X}_i(t) \big)
$$and thus there holds that  $\lim_{t\to \infty} \big\| \tilde{X}_i(t)- \frac{\sum_{i=1}^N X_i(0)}{N}\big\|=0$.

This completes the proof. \hfill$\square$

Combining Lemma \ref{lem3} and Proposition \ref{prop1}, Theorem \ref{theoremforgetting}.(i) is immediately  proved, where the exponential rate of convergence is simply resulted from the linear structure of the system (\ref{sysLind}).

\subsubsection{Diagonal Entries}
From the system (\ref{syscomponent}), we have
\begin{align}
\frac{d}{dt} \big[\rho(t) \big]_{|x\rangle \langle x|}= \sum_{\{j,k\}\in \mathsf{E}}  \Big(\big[\rho(t) \big]_{|u_{jk}(x)\rangle \langle u_{jk}(x)|}  - \big[\rho(t) \big]_{|x\rangle \langle x|}\Big),
\end{align}
which is consistent with the case when the network Hamiltonian $H$ is commutative with the swapping operators.  Theorem \ref{theoremforgetting}.(iii) readily follows from the analysis established in \cite{ShiDongPetersenJohansson} by applying Theorem 1 and Lemma 5 of \cite{ShiDongPetersenJohansson}.
\section{Conclusions}\label{Sec4}

We have made a further investigation to   the decoherence property of synchronization master equation for quantum networks of qubits interconnected  by swapping operators. The network Hamiltonian is assumed to be diagonal but with different diagonal entries so it might  not be commutative with the swapping operators. We proved  a theorem establishing a general  condition under which almost complete decohernece is achieved, i.e., all but two of the off-diagonal entries of the network density operator asymptotically tend to zero. This result explicitly revealed  that  quantum dissipation networks would  forget the information initially encoded when the internal (network Hamiltonian) and external (swapping operators) qubit interactions do not comply with each other. In future, it is interesting to look at the case with switching interactions where nontrivial coherence could be left if the switching signal properly responds to the network Hamiltonian.


\begin{thebibliography}{99}

\bibitem{tsi}
J. Tsitsiklis, D. Bertsekas, and M. Athans, ``Distributed asynchronous
deterministic and stochastic gradient optimization algorithms," {\em IEEE Trans. Autom. Control}, vol. 31, no. 9, pp. 803-812, 1986.

\bibitem{jad03}
A. Jadbabaie, J. Lin, and A. S. Morse,
``Coordination of groups of mobile autonomous agents using nearest neighbor rules,"
{\em IEEE Trans. Autom. Control}, vol. 48, no. 6, pp. 988-1001, 2003.



\bibitem{saber04} R. Olfati-Saber and R. M. Murray, ``Consensus problems in networks of agents with switching topology and time-delays," {\em IEEE Trans. Autom. Control}, vol. 49, pp. 1520-1533,  2004.

\bibitem{Magnus} M. Mesbahi and M. Egerstedt. {\em Graph Theoretic Methods in Multiagent Networks}. Princeton University Press. 2010.




\bibitem{Nielsen}  M. A. Nielsen, and I. L. Chuang. {\em Quantum Computation and Quantum Information}. 10th Edition. Cambridge University Press, 2010.







\bibitem{Sepulchre-non-communtative}
R. Sepulchre, A. Sarlette and P. Rouchon, ``Consensus in non-commutative spaces,"  {\em Proc. 49th IEEE Conference on Decision and Control}, pp. 6596-6601, Atlanta, USA, 15--17 Dec., 2010.

\bibitem{Ticozzi}  L. Mazzarella, A. Sarlette, and  F. Ticozzi, ``Consensus for quantum networks:
from symmetry to gossip iterations,"  {\em IEEE Trans. Autom. Control}, 60(1): 158--172, 2015.








\bibitem{Ticozzi-SIAM}  L. Mazzarella, F. Ticozzi and A. Sarlette, ``From consensus to robust randomized algorithms: A symmetrization approach," {\rm quant-ph, arXiv 1311.3364}, 2013.


\bibitem{ShiDongPetersenJohansson} G. Shi, D. Dong, I. R. Petersen, and K. H. Johansson, ``Reaching a quantum consensus: master equations that generate symmetrization and synchronization," {\it IEEE Transactions on Automatic Control}, in press,  {\rm arXiv:1403.6387}, 2015.

\bibitem{Altafini} C. Altafini,  ``Representing multiqubit unitary evolutions via Stokes tensors," {\em Physical Review A}, vol. 70, 032331, 2004.


%

\bibitem{NaturePhysiscs}F. Verstraete, M. M. Wolf, and J. I. Cirac, ``Quantum computation and quantum-state engineering driven by dissipation," {\it Nature Physics}, vol. 5,  pp.633--636, 2009.



\bibitem{Lindblad 1976}
G. Lindblad, ``On the generators of quantum dynamical semigroups,"
{\em Comm. Math. Phys.}, vol. 48, no. 2, pp. 119-130, 1976.

\bibitem{Breuer and Petruccione 2002}
H.-P. Breuer and F. Petruccione. {\em The Theory of Open Quantum
Systems}. Oxford University Press, 2002, 1st edn.



\bibitem{WU-CHUA95} C. W. Wu and L. O. Chua, ``Synchronization in an array of linearly
coupled dynamical systems,"  {\em IEEE Trans. Circuits
Syst.}, vol. 42, pp. 430-447, 1995.




\bibitem{WUbook} C. W. Wu. {\em  Synchronization in complex networks of nonlinear dynamical
systems}. World Scientific, 2007.

\bibitem{Dini} J. Danskin, ``The theory of max-min, with applications,"  {\em SIAM J. Appl. Math.}, vol. 14,
pp. 641-664, 1966.

\bibitem{Lin07} Z. Lin, B. Francis, and M. Maggiore, ``State agreement for continuous-time coupled nonlinear
systems," {\em SIAM J. Control Optim.}, vol. 46, pp. 288-307, 2007.

\bibitem{rou} N. Rouche, P. Habets, and M.
Laloy,
\newblock {\em Stability Theory by Liapunov's Direct Method}.
\newblock New York: Springer-Verlag, 1977.



\end{thebibliography}
\end{document}